# Memristors based Computation and Synthesis

Prashant Gupta
Dept. of Electrical and Computer Engineering
University of California, Davis
Email-pragupta@ucdavis.edu

Priscilla Jennifer
Dept. of Electrical and Computer Engineering
University of California, Davis
Email- pj@ucdavis.edu

*Abstract*— **Memristor has been identified as the fourth fundamental circuit element by Dr. Leon Chua in 1971 and since then it has gathered a lot of interest because of its non-volatility and thus are considered as a viable solution to the beyond CMOS era computation. Recently, Memristor have been used to perform basic logic operations like AND, OR, NAND, NOR etc. and appear to be useful for applications like Dot Product Engine, Convolution Neural Networks etc. This paper presents designing a Verilog behavioural model of Memristor and then using that to construct a 32-bit full adder. The paper later compares the Memristor with the current CMOS technology and highlights its advantages and pitfalls.**

## I. INTRODUCTION

Memristor (memory-resistor) is a two terminal fundamental passive element that was discovered by Dr. Leon Chua in 1971 as the fourth fundamental circuit element [1] besides the three well known circuit elements namely resistors, capacitors and inductors. For a long time the memristors remained just a theoretical idea until in 2008 when HP came up with an idea of fabricating nanoscale memristors using $TiO_2$ containing doped and un-doped regions. Fig.1, shows the memristor structure where **D** denotes the thickness of the sandwiched region and **w** is the thickness of the doped area (oxygen efficient area) in the $TiO_2$ memristor. Let $R_{ON}$ and $R_{OFF}$ denote the resistances at high and low dopant concentration areas, respectively.

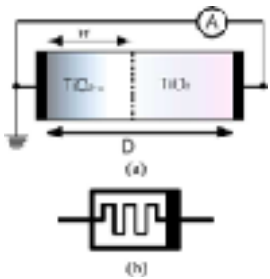

Fig. 1. (a) Structure of the $TiO_2$ memristor regions sandwiched between two platinum electrodes. When a voltage/current is applied, its memristance (resistance of the memristor in Ohms)/ memductance (conductance of the memristor)) is altered. (b) Symbol of the memristor. [2]

As the forward biased voltage is applied across the memristors the width of doped region increases (moves towards right) and hence, conductance of memristor increases as the resistance is reduced ($R_{ON}$). When a reverse biased voltage is applied across the memristor the width of doped region is reduced and hence the resistance of the memristor increases ($R_{OFF}$). In memristors the current voltage relationship is defined by the following equation-

$$v(t) = R(t)i(t) = (d\varphi/dq) * i(t) \quad (1)$$

where $\varphi(t)$ and $q(t)$ denote the flux and charge, respectively. A common well known characteristic of memristors and memristive devices is the pinched hysteresis loop in current vs voltage plane under sinusoidal excitations that goes through the centre. Due to this the resistance of the device depends on the past history of the input voltage or current.

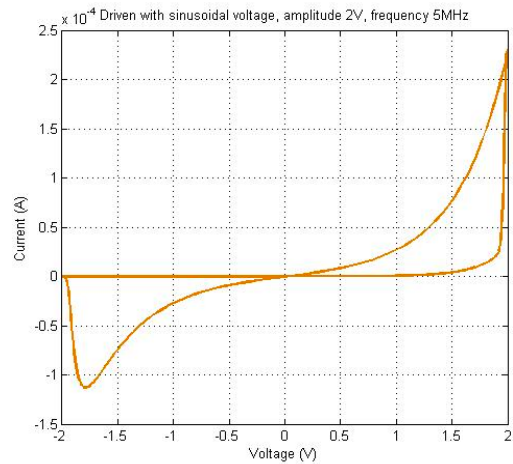

Fig. 2. The idealized memristive I-V characteristic shows the pinched hysteresis loop that passed through the o-region which shows that no energy is being stored in the memristors.

A lot of research groups have reported the use of memristors for basic boolean operations [3] [4] like AND, OR, NAND, NOR etc. which will be covered more in detail in Section III, but answering important architectural questions like scalability, power, area and timing analysis remains yet to be seen.

The end of CMOS and whats beyond CMOS is briefly discussed Section II, and Section III talks about key take aways of the existing work in the area of memristive logic synthesis, design styles, some high level applications using memristors and what's lacking in the current studies. Section IV, V and VI describes our memristor behavioural model and shows various logic simulation based on it. Section VII compares different memristor design styles with CMOS and discusses the advantages and disadvantages of using memristor over CMOS.

The paper is summarised in Section VIII with a discussion along with future research ideas and goals.

## II. BEYOND CMOS

Moore's law has served as a guiding principle for the semiconductor industry but the challenges have changed overtime due to data intensive application. Many factors mark the vitality of Moore's law from going forward [6]. The voltage scaling is one of the factors where the transistor's voltage cannot be further reduced due to leakage current in nanometer scale. Photolithography is another limitation to Moore's law as equipment and cost makes fabrication of transistors difficult for less than 7nm. The current focus of the semiconductor industry is to solve the concerns pertaining to scaling and power without compromising with the performance prompting researchers to explore new devices, architectures and technology to serve the emerging application like IoT (Internet of Things), cloud computing, wearables, real time data analytics etc. The target applications have changed over time and so the computational models have to updated to serve the applications demand.

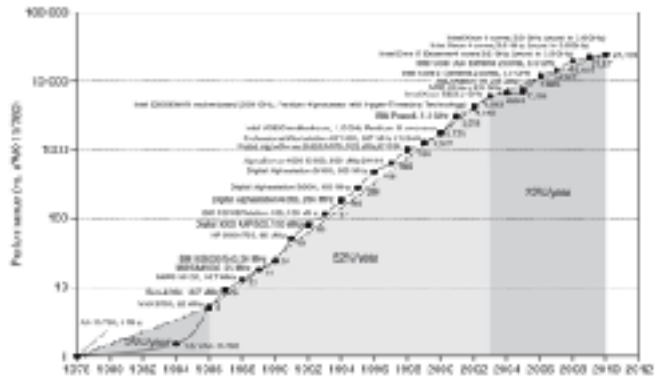

Fig. 3. Spectacular performance of CMOS coming to an end with lithographic equipment/cost limitations.

CMOS scaling has already reached the 14nm and scaling below 10nm makes it difficult to control properties of transistor to make it behave like an ideal switch [5]. The authors of International Technology Roadmap for semiconductors have challenged the computing and research community to find new devices and architectures that offer solution to memory and scaling bottleneck [4].

In 2008, a group of researchers in Hewlett Packard (HP) found a way to fabricate memristors which were first hypothesised in 1971. Memristors are dynamic devices with resistance that changes over time as a function of their existing state and voltages applied across them or the current driven through them.

Memristor, a nanoscale device can be used to perform stateful logic operations for which the same device can simultaneously serve as gates (logic) and latches (memory) which can be dynamically defined. They use resistance instead of voltage or current as the physical state variable and can be embedded with a nanoscale crossbar array to perform both store and logic using memory cells themselves. Another advantage of memristors is that they don't have to be constantly supplied with a voltage to retain the state information unlike CMOS where the state is lost once the supply voltage is removed. Memristors open an alternative solution to CMOS logic gates and give a viable solution to address the scaling and power bottleneck of transistor.

## III. EXISTING WORK

To understand memristor logic synthesis it is important to first understand the memristor design styles and what are the different ways to synthesise memristive logic gate. There are *three* popular design styles for memristor based circuits:

### A. IMPY Logic [4]

Implication logic or material implication or IMPLY logic, first described by Whitehead and Russel, is a fundamental Boolean logic operation on two variables p and q such that pIMPq ('p' implies 'q' ) means $(NOT p) OR q$. IMP and FALSE operations form a computationally complete logic basis i.e. we simulate any logic operation using combination of just these two operations. Figure 4, shows how material implication is naturally realized in a simple circuit combining a conventional resistor with two electrically connected memristor or memristive devices used as digital switches. The operation s←pNANDq can be implemented in a circuit with three memristive switches and P,Q and S. The computation is performed in three sequential steps:

$$s \leftarrow 0$$
$$s' \leftarrow pIMPs$$
$$s'' \leftarrow qIMPs'$$

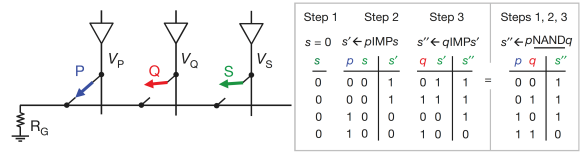

Fig. 4. The logic operation s←pNANDq performed as a sequential operation with three memristive switches. The voltages applied to obtain the NAND operation with P,Q and S with the truth tables showing the equivalence of the sequence of operations to NAND. [4].

The input values p and q stored in switches P and Q and the output was the value s'' accumulated in switch s. The selection of $R_G$, which may be the resistance of an appropriate length of a nanowire, is important to the outcome of the operation. The resistance is chosen such that $R_{CLOSED} < R_G < R_{OPEN}$ where, $R_{CLOSED}$ and $R_{OPEN}$ are the resistance states of the closed and open switches, respectively.

### B. MAGIC [3]

Memristor aided logic (MAGIC) is a method for memristor only logic. Like IMPLY logic, we can fabricate a crossbar using MAGIC enabling computing with memory but unlike IMPLY logic it does not require any resistors for implementing logic operations which makes the MAGIC implementation more efficient as it consumes less power and area. In MAGIC separate memristors are required for input and output. The inputs of the MAGIC gates are the initial logic state of the input memristors, and the output is the final resistance of the output memristor. Figure 5, shows MAGIC NOR implementation.

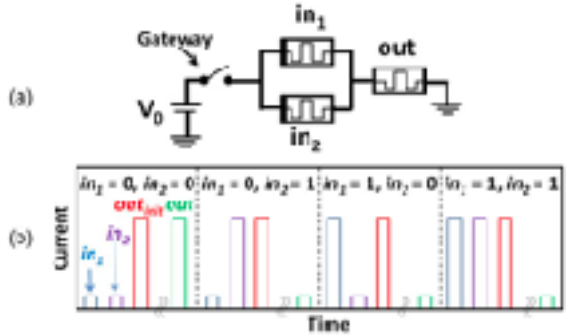

Fig. 5. a) Schematic of a two-input NOR gate. The logical gate consists of two input *in1* and *in2* and an out memristor *out*. During execution, a voltage $V_O$ is applied at the gateway of the circuit. b) Simulations of two-input NOR gate for all input combinations. Different curves show current reading from each memristor. [3]

The operation of MAGIC gate consists of two sequential stages. The first stage is initialising the output memristor to a known logic state (the initial state of output memristor is Logic 1 in case of gates like NOR and NAND while Logic 0 in case of gates like OR and AND). In the second stage of operation, a voltage $V_O$ is applied across the logic gate and the voltage across the output memristor is then evaluated. The voltage of the output memristor depends on the logic state of both the input memristors and initial state of the output memristor. Extending the logic to N inputs becomes much more simpler in case of MAGIC as we can simply stack up all the memristors to one single output memristor and if one of the those inputs will be Logic 1 then the output memristor flips its state.

C. *Hybrid Memristor-CMOS Logic* [7]

Hybrid Memristor CMOS Logic design is a fairly new idea and there are not enough studies available in this area. The basic understanding is that it contains both memristors and CMOS in their logic operations. As per the author in [7], IMPLY and MAGIC work fine standalone but become incompatible when trying to interface them with current generation CMOS process. So to integrate memristor with CMOS and to work with same voltage levels, there is a need of hybrid Memristor-CMOS. In this logic, the voltages are used as logic state. Memristors can only be used as computational element rather than computational cum storage element as used in material implication logic and MAGIC.

To realise a NAND gate using Hybrid Memristor-CMOS logic we use memristors to simulate AND logic and then CMOS inverter is used to invert the AND operation and realise the NAND gate. Figure 6, shows the NAND gate using Hybrid Memristor-CMOS logic operating at 1.8V.

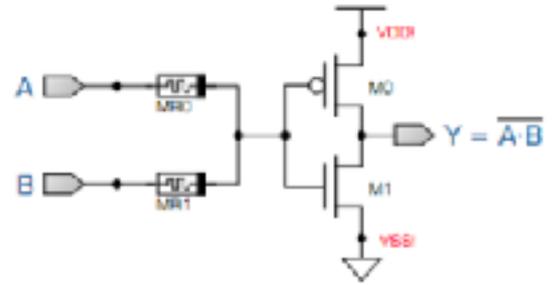

Fig. 6. Schematic of NAND gate using Hybrid Memristor-CMOS logic. The memristors are in the configuration to provide AND operation and CMOS NOT gate is used at the output to get NAND operation [7]

A lot of existing work in the field of memristors is based on designing memristive models and understanding their physical characteristics so as to fabricate better memristors. While HP in 2008 used $TiO_2$ [2] to first fabricate their memristor but some recent studies indicate that it might not be that efficient. With $TiO_2$ the width of the doped and un-doped region keeps on changing which leads to inconsistencies. Kristy A. Campbell, uses a permanent Ag conductive channel [8] and metal catalysed reaction within the device active layer is used to shift between the memristor logic states. The study shows that such a device can sustain high operating temperatures (150°C) and has faster switching.

HP in 2016 fabricated a 128x128 memristor crossbar and implemented a DOT product engine which does vector matrix multiplication[11]. The authors use a special conversion algorithm to map matrix values to crossbar array and claims 1000 times better performance than custom digital ASIC 512x512 crossbar array and can be used in neural network applications. The DOT product engine is later used in ISAAC architecture which is a convolutional neural network. The papers shows simulation and fabrication results of DOT product engine but information about implementation technique is vague which could have been critical for a study like this, where we try to answer architectural questions. The paper also makes significant assumptions in terms of power consumption where ADCs, DACs are considered as peripheral devices and the power consumed by them is not included in the overall system power.

Most of the memristor models present today for simulating memristors like Biolek [9], Joglekar, Knowm, Yakopcic, VTEAM [10] are either only to simulate only some physical characteristics of memristors or not well documented for anyone else to use. Hence, there was a need to create a new high level memristor behavioural model so that we could study memristor scalability and implement complex logic operations like 32-bit Full Adder which is covered in Section V.

## IV. MEMRISTOR MODEL

The memristor model proposed in this paper is high level Verilog behavioural model which can be used to build stateful logic gates and thus can also simulate complex combinational circuits like adders, multipliers etc.

The model matches the I-V characteristic of a memristor which can be considered similar to that of a CMOS latch that switches its state based on the input voltage and threshold voltage. In this model we assume memristor has a threshold of 0.5V.

************************************************

Memristor Model

************************************************

```verilog
1  //Behavioural Memristor Model
2  `celldefine
3  module memristor(vin,vout);
4  input vin;
5  output vout;
6  parameter vth = 0.5;
7  reg ps,vout;
8  always@(vin) begin
9  if (ps==0) begin
10  if(vin>vth)
11  ps=1;
12  end
13  else begin
14  if(vin<vth) begin
15  ps=0;
16  end
17  end
18  vout <=ps;
19  end
20  endmodule
21  `endcelldefine
```

************************************************

When the threshold voltage exceeds the input voltage, the memristor switches from Logic 0 to Logic 1 state and vice versa. The last value of the memristance of the memristor is retained as the output voltage. This output voltage actually stores the value of the updated state which is the characteristic behaviour of memristor. Another important criteria is initialisation of the states in memristors, the memristor has to be set to either logic 1 or 0 state depending on the logic operation. Thus for a NOR/NAND logic the output memristor state (ps) has to be initialised to logic 1 while for gates like AND/OR the output memristor state has to be initialised to logic 0.

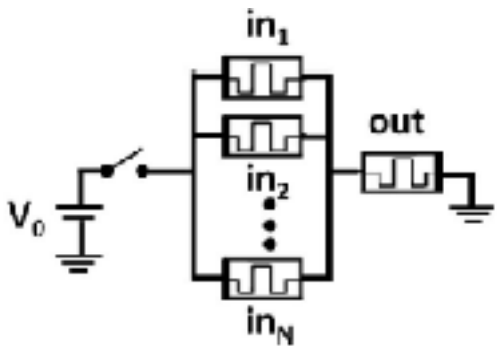

Fig.7. N-input NOR gate using MAGIC

For NOR operation, when both the input voltages are less than the threshold voltage (i.e. *In1* and *In2* are both Logic 0), the output memristor retains its state which is Logic 1. When one of the input voltage exceeds the threshold the logic state of output memristors switches from Logic 1 to Logic 0. This model can then be extended to N-input NOR which is shown as in the figure 7. In this paper we used the memristor behavioural model to implement a 16 input NOR gate and calculated the fan in of the gate.

## V. IMPLEMENTATION OF COMBINATIONAL CIRCUITS USING MEMRISTORS

### A. Full adder using MAGIC

The full-adder circuit adds three one-bit binary numbers (A,B, $C_{IN}$) and outputs two one-bit binary numbers, a sum (S) and a carry (C1). Here, with the NOR gate implemented from the model, we implemented a full adder using 9 NOR gates shown in figure 8. This architecture uses 9 NOR gates x 3 memristors = 27 memristors.

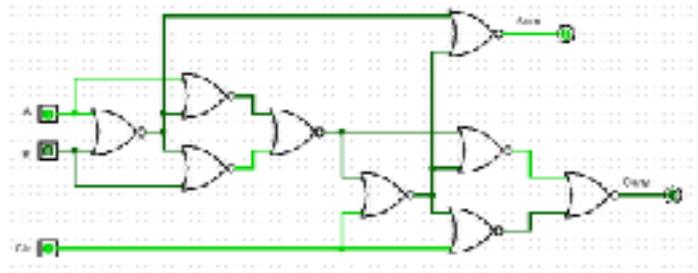

| Input | | | Output | |
|---|---|---|---|---|
| A | B | Cin | Sum | Carry |
| 0 | 0 | 0 | 0 | 0 |
| 0 | 0 | 1 | 1 | 0 |
| 0 | 1 | 0 | 1 | 0 |
| 0 | 1 | 1 | 0 | 1 |
| 1 | 0 | 0 | 1 | 0 |
| 1 | 0 | 1 | 0 | 1 |
| 1 | 1 | 0 | 0 | 1 |
| 1 | 1 | 1 | 1 | 1 |

Fig.8. Full Adder using NOR gates and its truth table.

### B. 32 Bit adder using MAGIC

Multiple full adder circuits can be cascaded in parallel to simulate a N-bit adder. For an N- bit parallel adder, there must be N number of full adder circuits. A ripple carry adder is a logic circuit in which the carry-out of each full adder is the carry in of the succeeding next most significant full adder. It is called a ripple carry adder because each carry bit gets rippled into the next stage. In a ripple carry adder the sum and carry out bits of any half adder stage is not valid until the carry in of that stage occurs. A 32 bit ripple carry adder using memristors is given in the figure 9.

The 32 bit ripple carry adder uses 32 full adder so the total number of memristor used in this circuit is 32 x 27 = 864 memristors, which is much less when compared to number of CMOS used in a 32 bit ripple carry adder which is, 32x 9 NOR gate x 4 transistors =1152 transistors. Hence, in 45nm CMOS technology for a 32 bit ripple carry adder 1152 transistors are required whereas only 864 memristors are needed when we use MAGIC.

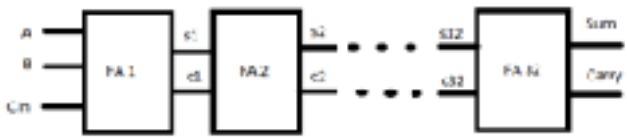

Fig.9. 32 bit Full Adder using cascaded 32 full adders

## VI. SIMULATION RESULTS

The verilog code is simulated using Cadence NCVerilog and SimVision simulator. The simulation results are obtained for memristors with a switching delay of 1ns between the logic states 0 and 1.

### 1) Simulation output for 2 input NOR

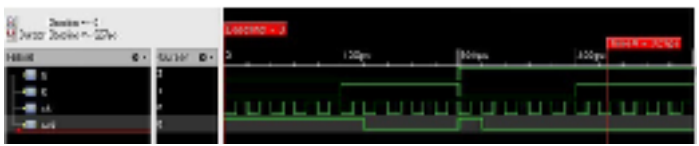

Fig.10. Simulation result for 2 input adder

### 2) Simulation output for 16 input NOR

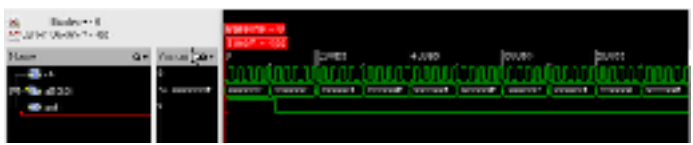

Fig.11. Simulation result for 16 input NOR

### 3) Simulation output for 2 input NOR driving 16 input NOR

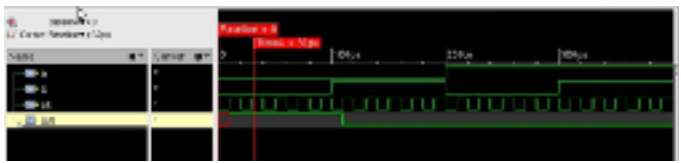

Fig.12. Simulation result for 2 input NOR driving 16 input NOR

### 4) Simulation output for full adder

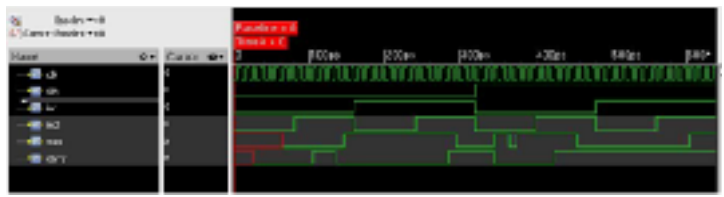

Fig.13. Simulation result for full adder

### 5) Simulation output for 32 bit adder

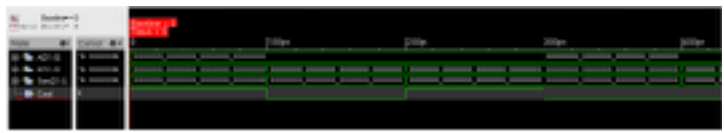

Fig.14. Simulation result for 32 bit adder

## VII. Logic Synthesis and Results

Logic Synthesis is the process of converting a high level description of design into an optimised gate-level representation. Logic Synthesis uses a standard cell library which has simple cells, such as basic logic gates like AND, OR, NOR, or macro cells, such as adder, mux, memory, and flip-flops. Standard cells put together are called technology library. Logic synthesis considers design constraints such as timing, area, testability, and power.

For our design we use 45 nm NanGate FreePDF45 Open Cell Library which is an open source library. We use Synopsys Design compiler for the synthesis of NOR gate, full adder and 32 bit adder and the results are tabulated in the Table I and Table II.

TABLE I. COMPARISON OF AREA AND SWITCHING POWER CHARACTERISTICS OF MEMRISTOR BASED CIRCUITS

|  | Area(um) | Switching Power (nW) |
|---|---|---|
| **2 Input NOR** | 0.05 | 46.87 |
| **16 Input NOR** | 0.28 | 46.87 |
| **2 Input NOR driving 16 Input NOR** | 0.33 | 46.87 |
| **Full Adder** | 0.45 | 93.75 |

TABLE II. COMPARISON OF 32-BIT ADDER USING CMOS AND MEMRISTIVE LOGIC

| 32 bit adder | Area(um) | Power(uW) | Time Delay (ns) |
|---|---|---|---|
| **CMOS** | 424.003991 | 100.5579 | 0.58 |
| **MEMRISTOR** | 289.407990 | 123.8836 | 3.63 |

Comparison of different parameters like area and clock switching power is given in Table I. The power consumed by 2 input NOR gate using CMOS transistors is less than 1uW.

Memristor based circuits are promising as they are very compact and consume less than 50nW power which could make it a potential replacement for CMOS solving problems like leakage power. In Table II , we compare the area, power and time delay of 32-Bit adder using CMOS Logic and Memristor MAGIC implementation. Here, the area of the adder using memristor is about half the area of adder using CMOS logic which could potentially solve the issue of CMOS scalability. As the memristors are passive circuit elements we would also need to consider power of different power lines for inputs and output for setting the initial conditions and the power boosting circuitry to compensate for the drop in voltage after every stage. Also, in this paper we use 45nm CMOS library for synthesis and thus, the results obtained are approximate values.

## VIII. DISCUSSION

This paper summarises existing work in the field of Memristors and then talks about our Memristor model, simulating logic gates and in the end compares memristors logic synthesis with CMOS. The results show thats we are successfully able to simulate 32-Bit adder using memristor NOR implementation.The results compare CMOS with memristor MAGIC implementation.

Given more time we would like to compare the results with IMPLY logic implementation. We would also like to make our memristor model more comprehensive by adding things like window function so that the states of the memristors changes with change in the width of doped and un-doped region and not just the input voltage. This will make the model more realistic to evaluate power of the entire circuit and not just the switching power.For adding window function to the model a lot of research need to be put into finding the right documentation so that models like Joglekar, VTEAM, Biolek etc. can be successfully integrated. As we are using 45nm CMOS library we could not benchmark it with realistic values and obtain exact results for digital logic circuits. Thus, having access to some memristor logic libraries for synthesising logic circuits and for timing analysis can prove extremely useful. It would also be interesting to evaluate memristors fabricated by Knowm Inc., to analyse the device properties so that we can better replicate it in the model.

The next step after the updated model could be using it in a crossbar array and then trying to simulate a Dot product engine with it. During our research we came across XbarSim tool [12] though which memristor crossbar based circuits can be simulated and tested with could also be useful. Understanding and developing the sequential circuit using memristors like state machines which could help in analysing the scope of memristor applications.

Studies have shown a variable switching behaviour below the threshold for the memristor elements [13]. Hence, leading to a probabilistic output for the logic gates. Building on this inherent feature of variability, the concept of approximate computing comes into place. The analog and mixed signal circuits are more prone to errors due to various factors like noise and device characteristics compared to digital logic circuits. Hence approximate computing finds application in this analog and mixed signal implementation. One such application is RRAM crossbar which uses memristor device in their crossbar arrays. This memristor crossbar uses analog matrix vector multiplication which is better in terms of performance and energy efficiency when compared to digital multiplication but suffers in terms of precision which is more suitable for approximate computing.

## ACKNOWLEDGEMENT

We will like to acknowledge Mr. Satyabrata Sarangi and Mr. Ajinkya More for their help throughout this project.